\documentclass[%
 reprint,
superscriptaddress,
 amsmath,amssymb,
 aps,
showkeys
]{revtex4-2}

\usepackage{tabularx}
\usepackage{graphicx}
\usepackage{dcolumn}
\usepackage{bm}
\usepackage{braket}
\usepackage[colorlinks, allcolors=blue]{hyperref}

\begin{document}


\title{Violation of Pauli Limit at KTaO$_{3}$(110) Interfaces} 
\author{Samuel J. Poage}
\altaffiliation{These authors contributed equally to this work}
\affiliation{Department of Materials Science and Engineering, The Ohio State University, Columbus, Ohio 43210 USA}
\author{Xueshi Gao}
\altaffiliation{These authors contributed equally to this work}
\affiliation{Department of Physics, The Ohio State University, Columbus, Ohio 43210 USA}
\author{Merve Baksi}
\altaffiliation{These authors contributed equally to this work}
\affiliation{Department of Physics, North Carolina State
University, Raleigh, North Carolina 27695, USA}
\author{Salva Salmani-Rezaie}
\affiliation{School of Applied and Engineering Physics, Cornell University, Ithaca, New York 14853, USA}
\affiliation{Kavli Institute at Cornell for Nanoscale Science, Cornell University, Ithaca, New York 14853, USA}
\author{David A. Muller}
\affiliation{School of Applied and Engineering Physics, Cornell University, Ithaca, New York 14853, USA}
\affiliation{Kavli Institute at Cornell for Nanoscale Science, Cornell University, Ithaca, New York 14853, USA}
\author{Divine P. Kumah}
\affiliation{Department of Physics, Duke University, Durham, North Carolina 27701, USA }
\author{Chun Ning Lau}
\affiliation{Department of Physics, The Ohio State University, Columbus, Ohio 43210 USA}
\author{Jos\'e Lorenzana}
\affiliation{ISC-CNR and Department of Physics, Sapienza University of Rome, Piazzale Aldo Moro 2, 00185, Rome, Italy}
\author{Maria N. Gastiasoro}
\affiliation{Donostia International Physics Center, 20018 Donostia-San Sebastian, Spain}
\author{Kaveh Ahadi}
 \email{ahadi.4@osu.edu}
\affiliation{Department of Materials Science and Engineering, The Ohio State University, Columbus, Ohio 43210 USA}
\affiliation{Department of Electrical and Computer engineering, The Ohio State University, Columbus, Ohio 43210 USA}

\date{\today}

\begin{abstract}
\section*{Abstract}
The superconducting order parameter at the KTaO$_{3}$ interfaces and its dependence on interface orientation remains a subject of debate. The superconductivity at these interfaces exhibits strong resilience against in-plane magnetic field and violates Pauli limit. The interface orientation dependence of critical field and violation of Pauli limit, however, have not been investigated. To address this problem, we grew epitaxial LaMnO$_{3}$/KTaO$_{3}$ heterostructures using molecular beam epitaxy. We show that superconductivity is extremely robust against the in-plane magnetic field. 
Our results indicate that the interface orientation, despite impacting the critical temperature, does not affect the ratio of critical field to the Pauli limiting field. These results offer opportunities to engineer superconductors which are resilient against magnetic field.
\end{abstract}

\maketitle


\section{Introduction}
Spin-orbit coupling (SOC) often enhances the resilience of Cooper pairs against magnetic field in two-dimensional (2D) superconductors \cite{saito2016highly, zhang2021ising}. Multiple SOC-related phenomena that could enhance critical field have been proposed including spin-orbit scattering \cite{WHH1966, klemm1975}, Rashba spin splitting \cite{gor2001superconducting}, and Ising pairing \cite{lu2015evidence, xi2016ising, falson2020type}.  

The emerging superconductivity at the KTaO$_{3}$ interfaces is sensitive to the crystallographic orientation with critical superconducting temperatures of $\sim0.05$ K \cite{ueno2011discovery}, $1$ K \cite{chen2021two}, and  $2$ K \cite{liu2021two} for (100), (110), and (111) interfaces, respectively. The superconducting order parameter and its dependence on interface orientation remains a subject of debate. The superconductivity at these interfaces exhibit strong resilience against in-plane magnetic field and violates Pauli limit \cite{liu2021two, al2023enhanced}. The interface orientation dependence of critical field and violation of Pauli limit, however, have not been investigated. 

Itinerant electrons occupy the tantalum 5$d$ derived $t_{2g}$ states with strong spin-orbit coupling in electron-doped samples \cite{himmetoglu2016transport}. To a good approximation, 
the  $t_{2g}$ states near $\Gamma$ can be described by an
$l=1$ orbital angular momentum\footnote{More precisely, the physical angular momentum of the $t_{2g}$ states can be mapped to minus the angular momentum of $p$ states so that ${\bm l}(t_{2g})=-{\bm l}(p)$. Since only ${\bm l}(p)$ satisfy canonical orbital momentum commutation rules mathematically, eigenstates are classified, defining an effective angular momentum operator, ${\bm j}_{eff}={\bm l}(p)+{\bm s}$. Neglecting crystal field effects in the ground state ($j_{eff}=3/2$) and   ${\bm l}$ and ${\bm s}$ are parallel. This should not be confused with the physical angular momentum 
${\bm l}(t_{2g})$ which is antiparallel to spin according to Hund's rule.  } which according to Hund's rule should be antiparallel to the spin. This results in a magnetic moment 
$\hat{\bm m} =-\mu_B (\hat{\bm l} +2 \hat{\bm s})\approx 0$ implying that quasiparticles are ``Hund's rule protected" from Zeeman coupling to magnetic fields leading to the possibility of overcoming the Pauli limit in superconductors. 
In surfaces, however, crystal fields may partially quench the angular momentum and break Hund's rule protection. We have already shown that this effect is small in (111) surfaces leading to a large violation of Pauli limit~\cite{al2023enhanced}. The (110) surface, however, has a lower symmetry which may also jeopardize Hund's rule protection through angular momentum quenching. We show that this protection is also strong in (110) oriented bilayers with crystal fields. 

Since KTaO$_{3}$ interfaces are not in the clean limit, we should also consider the effect of spin-orbit scattering. Indeed, in a system with sufficiently strong spin-orbit scattering rates the effects of spin paramagnetism are hampered, and the upper critical field can also greatly exceed the Pauli limiting field $H_P$~\cite{WHH1966,KLB1975}. We argue this is also a relevant mechanism to understand the extreme resilience against in-plane magnetic field in KTaO$_{3}$ heterostructures. 

Here we report on the emergence of 2D superconductivity with enhanced critical field at KTaO$_{3}$ (110) epitaxial interfaces. The in-plane critical field ($H_{c2}$) reaches $\sim 6$ T exceeding the Pauli field $H_P\approx 1.85$ T. KTaO$_{3}$ (111) interfaces reach in-plane $H_{c2}$ as high as $\sim 25$ T, which occurs at higher critical temperatures and carrier densities. The ratio of critical field to Pauli limit, however, remains independent of the interface orientation. Here, we compare the violation of Pauli limit at KTaO$_{3}$(110) and KTaO$_{3}$(111) interfaces. 

\begin{figure*}[t!]
\includegraphics[width=2\columnwidth]{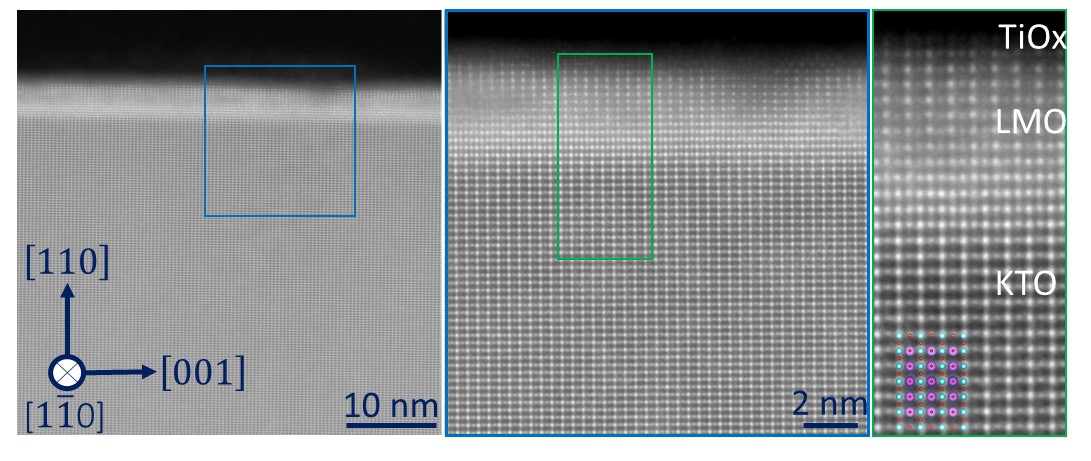}
 \caption{\textbf{The structure at the epitaxial LaMnO$_{3}$/KTaO$_{3}$(110) interfaces.} Cross-section HAADF-STEM images of LaMnO$_{3}$ film grown on KTaO$_{3}$ (110) substrate along [1$\bar1$0]  projection direction. }
    \label{fig:Fig1}
\end{figure*} 

\section{Results and Discussion}
\subsection{Experimental Results}
Epitaxial LaMnO$_{3}$/KTaO$_{3}$(110) heterostructures were grown in an oxide MBE. Figure S1 \cite{The} shows the reflection high-energy electron diffraction (RHEED) streaks, suggesting two-dimensional growth of LaMnO$_{3}$. High-angle annular dark-field scanning transmission electron microscope (HAADF-STEM) was used to investigate the grown heterostructures. Figure 1 shows the HAADF-STEM image of LaMnO$_{3}$ grown on a KTaO$_{3}$(110) substrate along [001] projection direction. Figure S2 \cite{The} shows the HAADF-STEM of the same sample along [1$\bar1$0] projection direction. The HAADF-STEM image confirms epitaxial growth of LaMnO$_{3}$ on KTaO$_{3}$(110). The regions that appear rougher in LaMnO$_{3}$ could be due to the steps at the (110) interface.

Figure S3 \cite{The} exhibits the Hall carrier density with temperature. The carrier density is $n_{s}=6.9\times10^{13}$ $cm^{-2}$ at 2 K. The freeze out of charge carriers could be due to defect states, including oxygen vacancies, acting as traps \cite{ojha2021oxygen}. The measured 2D carrier density is within the range that a superconducting transition is expected \cite{chen2021two}. 

\begin{figure}[]
    \centering
    \includegraphics[width=1.\columnwidth]{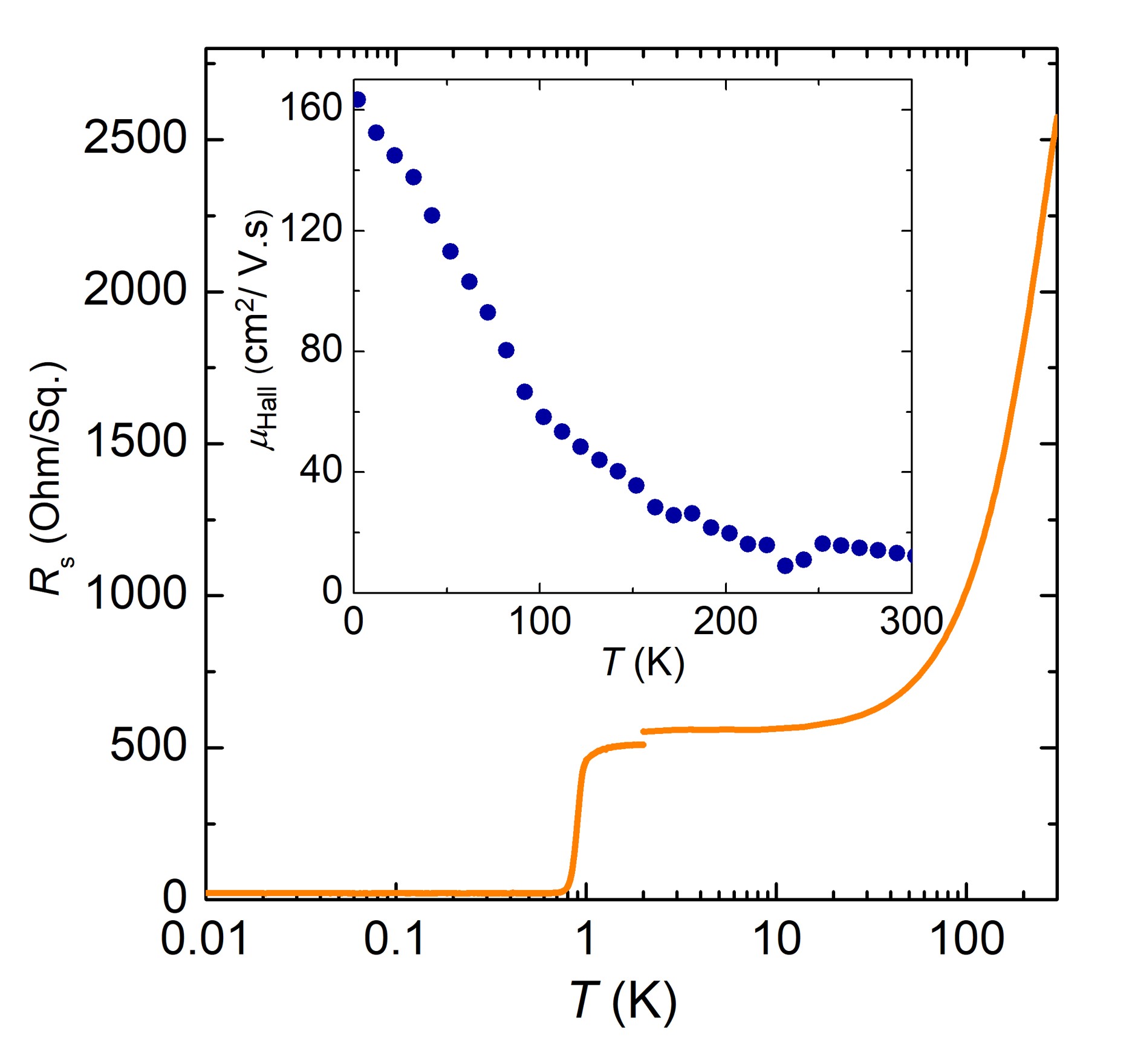}
 \caption{\textbf{Charge transport and superconducting transition at the LaMnO$_{3}$/KTaO$_{3}$(110) interfaces.} Sheet resistance with temperature. Sample exhibits metallic behavior, $dR/dT>0$, followed by an abrupt superconducting transition at $\sim1$ K. Inset exhibits Hall carrier mobility with temperature, $\sim160$ $cm^{2}/v.s$ at 2 K. High temperature (300-2 K) and low temperature (2-0.01 K) measurements were carried out in a Physical Property Measurement System and dilution refrigerator, respectively. }
    \label{fig:Fig2}
\end{figure}

Figure 2 shows the sheet resistance with temperature (300-0.1 K), measured along the [001] direction. All films exhibit a metallic behavior, $dR/dT>0$, followed by an abrupt superconducting transition. Heterostructures, grown on sapphire substrate instead of KTaO$_{3}$, show insulating behavior, suggesting the transport occurs at the KTaO$_{3}$ side of the interface.

Figure S4 \cite{The} shows the superconducting transition with temperature measured along [001] and [1$\bar1$0] directions in various samples grown with similar conditions. All samples have somewhat similar critical temperatures of superconductivity $\sim1$ K (defined at 0.8 $R_n$), which is comparable to those found at EuO/KTaO$_{3}$(110) and LaAlO$_{3}$/KTaO$_{3}$(110) interfaces \cite{chen2021two, hua2022tunable}. The superconducting transition, however, is slightly different in some of the samples along [001] and [1$\bar1$0] directions. Strong anisotropic superconductivity was previously reported in  KTaO$_{3}$(111) interfaces \cite{liu2021two} and was attributed to a stripe electron fluid \cite{villar2021}, a nematic state \cite{buessen2021nematic} and the anisotropy of density of states at Fermi surface \cite{arnault2023}. Furthermore, spontaneous symmetry breaking transitions were reported at KTaO$_{3}$ interfaces at low temperature in anisotropic magnetoresistance (AMR) \cite{wadehra2020planar} and near the critical temperature of superconductivity \cite{zhang2023spontaneous}. 

The  KTaO$_{3}(110)$ surface has two-fold rotational symmetry. As a consequence, the electronic structure is expected to be anisotropic \cite{martinez2023anisotropic}. Figure S5 \cite{The} shows the AMR measured at 1.7 K and various magnetic fields. Here, AMR results exhibit a trivial two-fold symmetric response and no sign of symmetry breaking transition up to 14 T, similar to SrTiO$_{3}$ \cite{miao2016anisotropic} and EuTiO$_{3}$ \cite{ahadi2019anisotropic} at higher fields. 

The inset in Figure 2 shows the Hall carrier mobility with temperature reaching $\sim160$ cm$^{2}$/(V sec) at 2~K, in line with high quality KTaO$_{3}$ 2D electron systems \cite{liu2023tunable, arnault2023}. The mean free path of charge carriers, assuming a single-band model, is $\approx 22$ nm, $l_\mathrm{mfp}=\frac{h}{e^2}\frac{1}{K_f R_s}$, where $K_f=\sqrt{2\pi n_s}$. From the low temperature carrier mobility, we estimated the relaxation times as $\tau_\mathrm{tr}=m^* \mu/e$, assuming the effective mass $m^*=0.23$\cite{himmetoglu2016transport}. We obtain $\tau_\mathrm{tr}=2.1 \times 10^{-14}$~sec. The BCS superconducting gap ($\Delta \approx 1.76 k_B T_c$) is 152~$\mu e$V, yielding $\Delta \tau_\mathrm{tr}/\hbar=0.005$ well in the dirty regime. This is similar to what was found in the (111) interface 
$\Delta \tau_\mathrm{tr}/\hbar=0.003$ and calls for the analysis of the effect of spin-orbit scattering. 

\begin{figure}[]
    \centering
    \includegraphics[width=0.8\columnwidth]{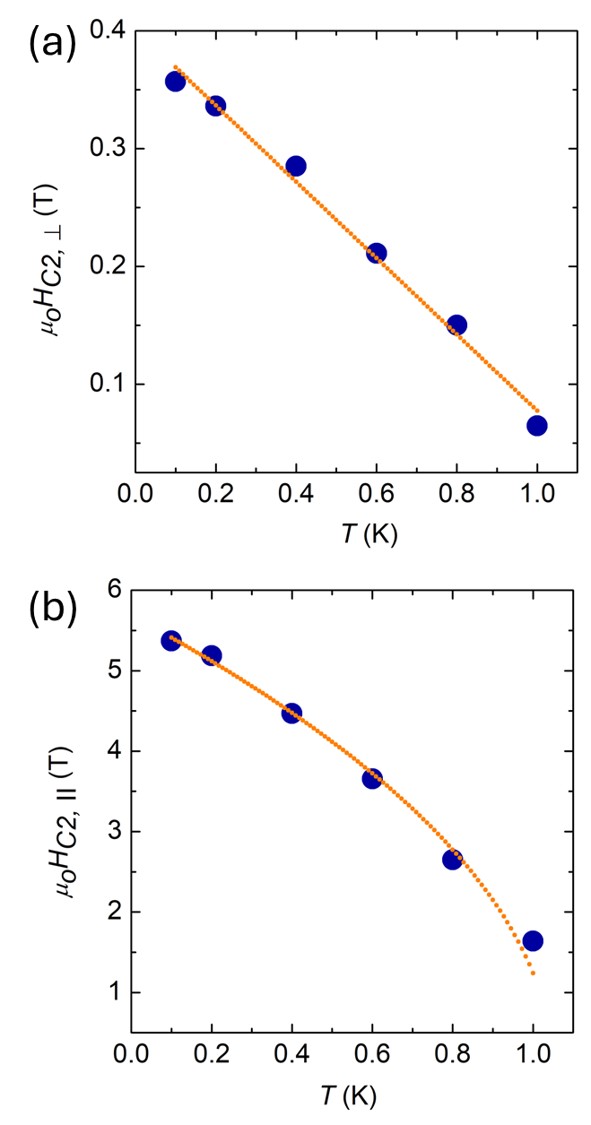}
 \caption{\textbf{Critical field of superconductivity with temperature at the LaMnO$_{3}$/KTaO$_{3}$(110) interfaces.} (a) out-of-plane and (b) in-plane critical fields of superconductivity, defined at 0.8$R_{n}$, with temperature measured along [001] direction. The orange line is a fit to measured data. The out-of-plane critical field with temperature is fitted using a Ginzburg-Landau (GL) model, $H_{c2,\bot}=\Phi_0(1-T/T_{c})/[2\pi(\zeta_{GL})^2]$ where $\Phi_0$ and $\zeta_{GL}$ are magnetic flux quantum and coherence length, respectively. The GL coherence lenght is $\zeta_{GL}$$\approx 29$ nm. The in-plane critical field is described using a square root behavior\cite{tinkham1963effect}, $H_{c2,\parallel}=\Phi_0[12(1-T/T_{c})]^{0.5}/[2\pi(d_{sc}\zeta_{GL})]$, where $d_{sc}$ is the effective thickness of superconducting layer, estimated to be $\sim7$ nm. }
    \label{fig:Fig3}
\end{figure}

We further study the superconducting transition by measuring the longitudinal transport with applied magnetic field. Figure S6(a) and (b) \cite{The} show the longitudinal magnetoresistance at various temperatures with in-plane and out-of-plane applied magnetic field. Figure 3 shows the out-of-plane critical field ($H_{c2,\bot}$) and in-plane critical field ($H_{c2,\parallel}$), defined at 0.8$R_{n}$, with temperature. The critical field monotonically increases as the sample is cooled. The critical field defined at other normal state resistance fractions also follow a similar trend (see Figure S6(c) and (d)).

Figure S7(a) \cite{The} exhibits the superconducting transition with magnetic field applied at various angles with respect to current. The critical field increases as the angle between applied field and film reduces, suggesting a significant asymmetry between out-of-plane and in-plane $H_{c2}$, an effect commonly observed in 2D superconductors \cite{saito2016highly}. Figure S7(b) \cite{The} shows the anisotropic critical field ratio ($H_{c2,\parallel}/H_{c2,\bot}$) remains $\sim15$ at various temperatures.

The out-of-plane critical field with temperature can be fitted by the Ginzburg-Landau (GL) theory, $H_{c2,\bot}=\Phi_0(1-T/T_{c})/[2\pi(\zeta_{GL})^2]$ where $\Phi_0$ and $\zeta_{GL}$ are magnetic flux quantum and coherence length, respectively, giving an extrapolated $H_{c2,\bot}(0)$=0.4 T. The GL fit yields $\zeta_{GL}$$\approx 29$ nm. The in-plane critical field with temperature follows a square root behavior, introduced by Tinkham \cite{tinkham1963effect}, $H_{c2,\parallel}=H_{c2,\parallel}(0) [(1-T/T_{c})]^{0.5}$.  The extrapolated parallel critical field from this fit is $H_{c2,\parallel}(0)$=5.7~T. 

Two different mechanisms may determine $H_{c2,\parallel}(0)$. In the case of orbital currents determining the critical field, a GL estimate  yields
$H_{c2,\parallel}(0)=\Phi_0 \sqrt{3}/(\pi d_{sc}\zeta_{GL})$  where $d_{sc}$ is the effective thickness of the superconducting layer.  Using the above values one obtains $d_{sc}\sim7$ nm.

Alternatively, superconductivity may be quenched at a first-order transition when the Pauli paramagnetic limit is reached. Assuming weak coupling superconductivity and equating the normal state and superconducting free energies one obtains that the critical field is given by, 
\begin{equation}
    H_{c2,\parallel}= H_P \sqrt{\frac{\chi_P}{\chi_N-\chi_{SC}}},\label{eq:pauli} 
\end{equation}
where $H_P=\Delta_0/\sqrt{2}$ is the Pauli field, $\chi_P=\mu_B^22N_F$ the Pauli spin susceptibility of non-interacting electrons with density of states per spin at the Fermi level $N_F$, and $\chi_{N(SC)}$ is the normal-state (superconducting) magnetic susceptibility. For singlet superconductivity and in the absence of spin-orbit coupling 
$\chi_{SC}=0$,  $\chi_N=\chi_P$ and
$H_{c2,\parallel}(0)=H_p=  \frac{\Delta}{\sqrt{2}\mu_B} \approx1.85$~T.

This means that the in-plane critical field considerably exceeds the Pauli limit in this system, $H_{c2,\parallel}/H_p$=3.1. Violation of Pauli limit was previously reported at KTaO$_{3}$ (111) interfaces in which the $H_{c2,\parallel}/H_p$ ratio increases with carrier density, reaching a ratio of 8 for $n_{s}=9\times10^{13}$ cm$^{-2}$. Figure S8 \cite{The} exhibits the $H_{c2,\parallel}/H_p$ ratio with carrier density for KTaO$_{3}$(111) and (110) interfaces. Interestingly, the $H_{c2,\parallel}/H_p$ ratios in KTaO$_{3}$(111) and (110) interfaces are similar and only depend on the carrier density. This indicates that despite the strong dependence of critical temperature on interface orientation, the critical field scales linearly with $T_c$ between KTaO$_{3}$(111) and (110) interfaces, keeping the $H_{c2,\parallel}/H_p$ ratio unchanged.

\subsection{Mechanism for critical field enhancement} 

One often-quoted mechanism for an enhanced critical field is unconventional pairing. For example, Pauli limit does not apply to $p$-wave superconductors and enhancement of in-plane critical field with mixed parity has been reported in SrTiO$_{3}$ \cite{schumann2020possible}. Signatures of mixed-parity superconductivity, however, have not been realized at KTaO$_{3}$ interfaces \cite{arnault2023}. Fulde–Ferrell–Larkin–Ovchinnikov (FFLO) state could also enhance critical field \cite{fulde1964superconductivity}. The FFLO state emerges in the clean regime $\Delta \tau_\mathrm{tr}/\hbar\gg 1$ whereas here, as discussed above,  the system is deep in the dirty regime. 

Previously\cite{al2023enhanced} we have identified two distinct mechanisms that can contribute to overcoming the Pauli limit in KTaO$_{3}$ heterostructures: Hund's rule protection and strong spin-orbit scattering due to quenched disorder. The former involves quasiparticles with reduced magnetic moments and correspondingly small Zeeman coupling to magnetic fields.
The latter randomizes the spins, screening the polarizing effect of the applied magnetic field ~\cite{WHH1966,Hwang2012,Hwang2018,lu2014}.

We first consider Hund's rule protection in the clean limit. According to Eq.~\eqref{eq:pauli} the Pauli limit can be overcome by an admixture with triplet superconductivity leading to $\chi_{SC}>0$ or by a reduction of the normal state magnetic susceptibility due to spin-orbit effects, $\chi_{N}<\chi_{P}$~\cite{yip2013,Xie2020,al2023enhanced}. 

Since inversion symmetry is broken at the interface, an admixture of triplet superconductivity could be present. This will contribute\cite{gor2001superconducting} a maximum critical field enhancement with respect to $H_P$ of  $\sqrt2$ which is too small to explain the factor of $\sim 3.1 $  observed. Therefore, in the following we neglect this effect and assume  $\chi_{SC}=0$, hence our estimates will be a lower bound for the enhancement of in-plane $H_{c2}$.

\begin{figure*}[]
    \centering
    \includegraphics[width=2\columnwidth]{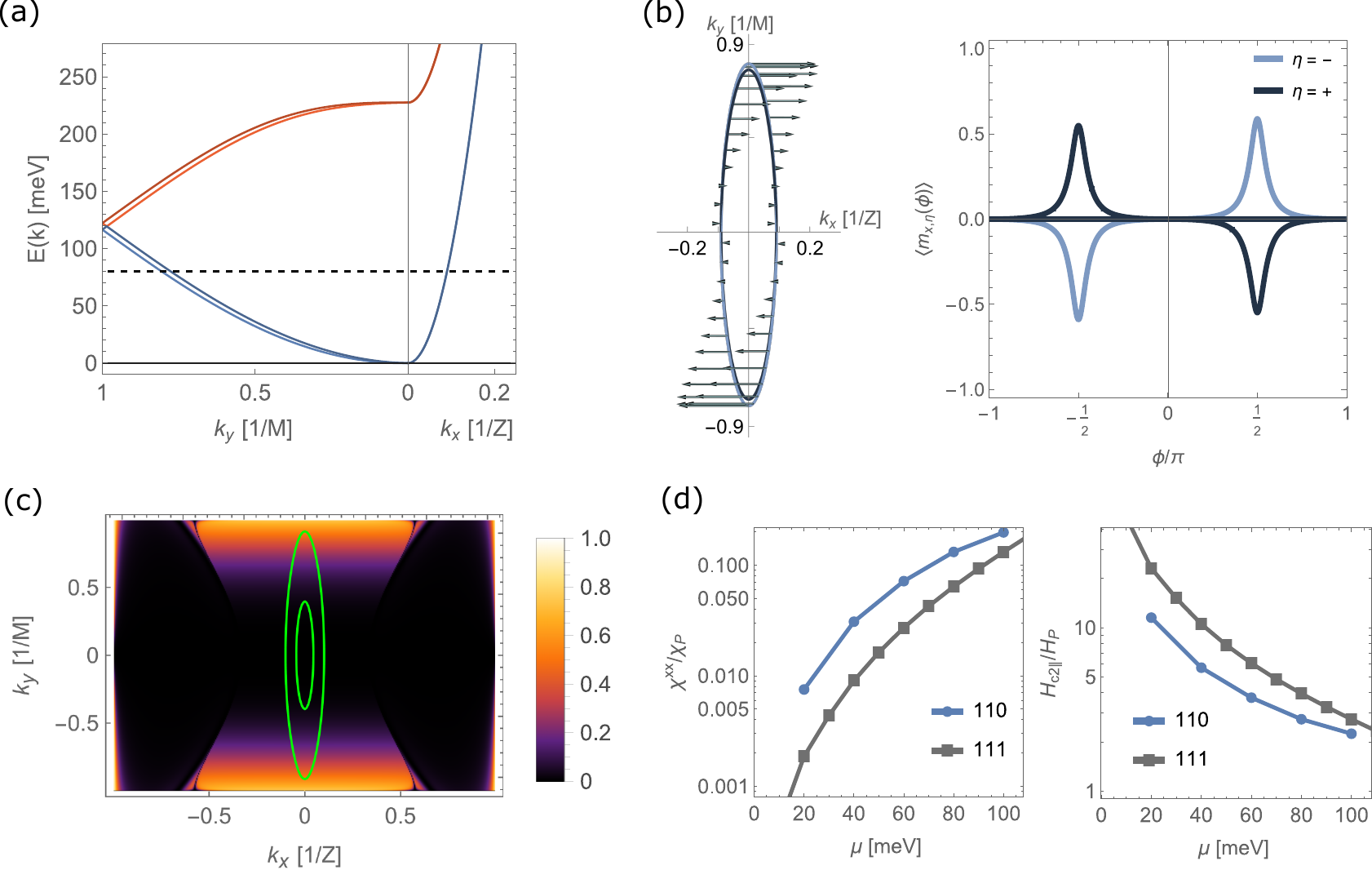}
 \caption {\textbf{Hund's rule protection in the (110) bilayer model. } (a) Electronic band structure of the 110 bilayer model [see SM] along two high symmetry in-plane momenta $k_x$ and $k_y$ in units of $Z=\frac{\pi}{a}[001]$ and $M=\frac{\pi}{\sqrt{2}a}[1\overline 1 0]$, respectively. A finite Rashba coupling $t_\mathrm{R}=2$ meV splits the double degeneracy of the lower band $n=1$ (blue) and upper band $n=2$ (red) into opposite helicities $E_{n\pm}(k)$. (b) The Fermi surface for $\mu=80$ meV (dashed black line in (a)) and magnetic moment texture (shown only for outer helicity $\eta=-$ in left panel for clarity) is parallel to $k_x$ ($\langle m_{y/z}\rangle\rightarrow 0$) and significantly reduced from 1, particularly close to the $k_x$ axis ($\phi\rightarrow 0,\pm\pi$). (c) Matrix elements $\gamma^x_{1-}(\mathbf{k})$. The green contours show the Rashba band $E_{1-}(\mathbf{k})$ Fermi surface for $\mu=20$ meV (inner contour) and $\mu=100$ meV (outer contour). (d) Spin susceptibility $\chi^{xx}$ of the 110 bilayer (blue) and 111 bilayer~\cite{al2023enhanced} (gray) normalized to the Pauli susceptibility $\chi_{P}$ (left) and corresponding $H_{c2,\parallel}$ in units of the Pauli-limit critical field (right) as a function of the chemical potential. }
    \label{fig:Fig4}
\end{figure*}

To proceed we need to compute the normal state susceptibility in the presence of spin-orbit coupling for the heterostructure. We model the (110) surface with a tight binding bilayer model~\cite{villar2021,liu2023tunable} and the same parameters that we used before for the (111) surface (see SI for details). Figure 4(a) shows the band structure of the bilayer model. There is a strong mass anisotropy so the dispersion is much steeper along $x$ ([001]) than along $y$ $([1\overline10])$. Accordingly, the Fermi surface is very elongated as shown in Figure 4(b). Notice the Rashba splitting of bands with opposite helicity along the weakly dispersive direction. 

Figure 4(b) also shows the magnetic moment,  $\braket{\bm{m}_\eta}(\bm{k})=\bra{n\eta\bm{k}}(\bm{\hat l} +2 \bm{ \hat s})\ket{n\eta\bm{k} }$ for quasiparticles of the lowest band $n=1$, helicity $\eta=-$, and momentum $\bm{k}$ on the Fermi surface. 
Due to the anisotropy of the electronic structure, the moment is oriented along $x$ along the entire Fermi surface. The orbital angular momentum $\langle\bm{\hat l}\rangle$ is nearly 1 close to the zone center (where SOC coupling dominates) and antiparallel to $\langle\bm{\hat s}\rangle$. As a result, the magnetic moments of the quasiparticles are quite small along most of the Fermi surface with the largest moment $<0.6 \mu_B$ in the point further away from $\Gamma$ (along $k_y$). A finite crystal field further splits the $\eta=\pm$ Rashba bands, decreasing (increasing) the magnetic moment of the inner (outer) band. Hence, its overall effect on the Zeeman coupling is very small and we neglect it from now on for simplicity.  

Here, we focus on the case of a magnetic field along the $x$ direction, as the case of a magnetic field along $y$ needs a model with more states leading to a finite component of the magnetic moment. 
The normal-state in-plane magnetic susceptibility is $\chi_n^{xx}=g^2\mu_B^2 N_{F,n}/2$, where the effective $g$-factor is given by $g^2=2\sum_{\eta=\pm} \gamma^x_{n\eta,F}$,
 $\gamma^x_{n\eta,F}$ is the Fermi surface average of the matrix element squared, 
$|\braket{m_{x,\eta}(\bm{k})}|^2$, and $N_{F,n}$ is the density of states \emph{per spin} at the Fermi level of band $n$. In the absence of SOC, $\gamma^x_{n\eta,F}=1$ and we recover the Pauli susceptibility $\chi^{xx}_n=\chi_{P,n}=\mu_B^2 2N_{F,n}$.  
Figure 4(c) shows the matrix element $|\braket{m_{x,-}(\bm{k})}|^2$
plotted in the Brillouin zone superimposed to two Fermi surfaces for different chemical potential values $\mu$. As seen, in most of the Fermi surface the matrix element is suppressed (smaller than 1) and this is more so for small $\mu$. Accordingly, the normal state susceptibility is strongly reduced for small chemical potential and the critical field is much larger than the Pauli field [Figure 4(d), blue lines]. The magnitude of the enhancement is similar to what was obtained for the (111) bilayer model with similar parameters (gray lines).

Experimentally, the surpass of the Pauli limit increases with doping while in the present mechanism it decreases with higher chemical potential. Considering the effect of disorder, the trend is inverted\cite{al2023enhanced}. Indeed, strong spin-orbit scattering can also lead to an enhancement of the critical field. In the Klemm–Luther–Beasley (KLB) theory~\cite{KLB1975}, appropriate for systems in the dirty limit with strong SOC scattering, the critical field $H_c$ is given by
\begin{equation}
\label{eq:klb}
    \ln\left(\frac{T}{T_c}\right)=\psi\left(\frac{1}{2}\right)-\psi\left(\frac{1}{2}+\frac{3\tau_\mathrm{SO}}{2\hbar}\frac{g^2\mu_B^2H_c^2}{8\pi\kappa_B T}\right)
\end{equation}
with the spin-orbit scattering time $\tau_\mathrm{SO}$ and the digamma function $\psi(x)$. In this equation, Hund's rule protection enters through the effective $g$-factor.

\begin{figure}[]
    \centering
    \includegraphics[width=0.7\columnwidth]{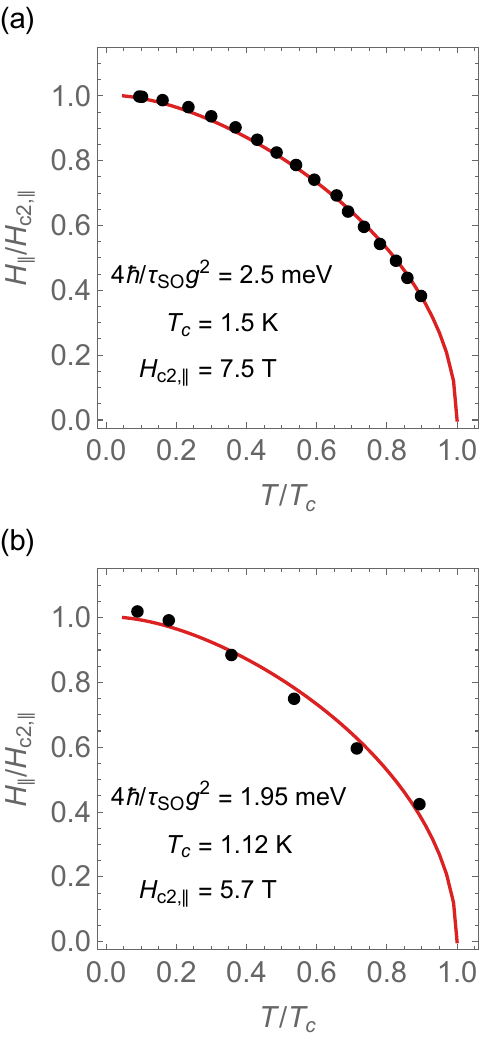}
 \caption{\textbf{Critical field in the presence spin-orbit scattering. } Klemm–Luther–Beasley (KLB) fit of the in-plane critical field  [Eq.~\eqref{eq:klb}], defined at 0.9$R_{n}$, with temperature at the (a) LaMnO$_{3}$/KTaO$_{3}$(111) (b) LaMnO$_{3}$/KTaO$_{3}$(110) interfaces.  The critical fields with temperature exhibit a monotonic increase of $H_{c2}$ as we cool the samples. The LaMnO$_{3}$/KTaO$_{3}$(111) and LaMnO$_{3}$/KTaO$_{3}$(110) samples have Hall carrier densities of $n_{s}=6\times10^{13}$ $cm^{-2}$ and $6.9\times10^{13}$ $cm^{-2}$ at 2 K, respectively. }
    \label{fig:Fig5}
\end{figure} 

Figure 5 exhibits a fit to the experimental data for KTaO$_{3}$ (111) and (110) interfaces with similar carrier densities. We use $g^2\tau_{SO}$ as a fitting parameter. Unfortunately, it is not possible from the present measurements to obtain independent values of $g^2$ and $\tau_{SO}$. Therefore, we can not disentangle the impact of the two mechanisms on increasing the critical field beyond $H_P$.  

For the present (110) interface we obtain $g^2\tau_\mathrm{SO}= 3 \times 10^{-13} \sec $. Independently of the value assumed for $g^2$ (assuming it is smaller than the free electron value, $g^2=4$) we find  $\tau_\mathrm{SO}\gg 
\tau_\mathrm{tr}$, as expected. This clearly points to an important role of the SOC scattering in the mechanism for enhancing the in-plane critical fields in KTaO$_{3}$ heterostructures.

\section*{Conclusions}

In summary, our experimental results demonstrate very large in-plane critical fields $H_{c2, \parallel}$ well beyond the Pauli limit. Our results also indicate that the extent to which the Pauli limit is surpassed, the $H_{c2,\parallel}/H_p$ ratio, only depends on the carrier density and remains unchanged between KTaO$_{3}$(111) and (110) interfaces. 

Theoretically, we have discussed two mechanisms that can enhance the critical field beyond $H_P$. Hund's rule protection enters through the effective $g$-factor which determines the Zeeman coupling. We have presented a simplified bilayer model of the band structure to study this effect. We find that the ratio of the critical field to the Pauli field depends weakly on the orientation of the interface, which is consistent with the experiments. In the future, it would be interesting to study both theoretically and experimentally the in-plane anisotropy of the enhancement in the (110) interface. Theoretically, this requires a more realistic model of the electronic structure to avoid unphysical features given by the present setting.

Our results also show that spin-orbit scattering is essential to describe the enhancement of the critical field, as it restores the observed doping dependence. As a byproduct, our computations predict that in very clean samples, the doping dependence of the enhancement should be inverted. It would be interesting to test this intriguing possibility in a field effect transistor configuration.

\section*{Data Availability Statement}
 
The data that support the findings of this study are available in the article and its Supplemental Material. Raw data can be obtained from the corresponding authors upon request.

\begin{acknowledgments}
OSU team was supported by the U.S. National Science Foundation under Grant No. NSF DMR-2408890. The microscopy effort is based upon work supported by the National Science Foundation (Platform for the Accelerated Realization, Analysis, and Discovery of Interface Materials (PARADIM)) under Cooperative Agreement No. DMR-2039380. This work made use of a Helios FIB supported by NSF (Grant No. DMR-1539918) and the Cornell Center for Materials Research (CCMR) Shared Facilities, which are supported through the NSF MRSEC Program (Grant No. DMR-1719875).  M.N.G is supported by the Ramon y Cajal Fellowship RYC2021-031639-I funded by MCIN/AEI/ 10.13039/501100011033.  
\end{acknowledgments}

\section*{Appendix: Materials and methods}

Heterostructures were grown in an oxide MBE with base pressure of $2\times10^{-10}$ Torr. The carrier density was controlled with growth parameters, described elsewhere \cite{al2022superconductivity, al2021two, arnault2023}. 10 u.c.s of LaMnO$_{3}$ was grown followed by TiO$_{x}$ (1-3 nm) on KTaO$_{3}$ (111) substrate. The elemental fluxes, emanating from effusion cells, were calibrated using a Quartz Crystal Microbalance (QCM). 

Substrate was annealed for 30 min prior to the growth at 600 $ ^\circ$C and, subsequently, heated to 800 $ ^\circ$C and growth started immediately to avoid potassium loss. Oxygen partial pressure was $3\times10^{-6}$ Torr during the LaMnO$_{3}$ growth. Reflection high-energy electron diffraction (RHEED) shows diffraction streaks, suggesting smooth LaMnO$_{3}$ films (Fig. S1). The samples were cooled to 550 $ ^\circ$C, immediately after the growth of LaMnO$_{3}$, to grow TiO$_{x}$ amorphous layer ($\sim$3 nm). Disappearance of RHEED streaks confirms amorphous nature of TiO$_{x}$ layer. Finally, a 20 nm amorphous lanthanum oxide layer was grown as a capping layer, protecting the 2DEG from oxidation. The growth was repeated on sapphire substrate, which resulted in an insulating behavior ($\sim$1-10 MOhms).

Scanning transmission electron microscopy (STEM) was performed to study the films and their interface structures. Cross-section samples were prepared using a Thermo Fisher Scientific Helios G4UX focused ion beam. High-angle annular dark-field (HAADF) images were obtained using a Thermo Fisher Scientific Spectra 300 X-CFEG operating at 200 kV with a convergence angle of 30 mrad and a HAADF detector with an angular range of 60-200 mrad. 

The high temperature (300-3 K) magneto-electric measurements were carried out in a Quantum Design Physical Property Measurement System (PPMS). Transport measurements were carried out in van der Pauw geometry with square-shaped samples and gold contacts deposited on the sample corners using a sputtering system. The Hall coefficient was calculated from a linear fit to R$_{xy}$ with magnetic field ($R_{H}$=$d$R$_{xy}$/$d$B).The Hall carrier density was extracted from 
$n_{2D}=-1/(eR_{H}$), where $e$ is the electron charge. 

The sub-Kelvin magnetoelectric measurements were carried out in an Oxford dilution refrigerator with base temperature of 10 mK as measured by Cernox sensors (50 mK with the application of high field). This refrigerator is equipped with a 16T superconducting magnet, which provides the magnetic field in the experiments. All electrical lines are fitted with QDevil filters, which comprises of 3-stage low frequency RC filters. Longitudinal resistance R$_{xx}$ and transverse resistance R$_{xy}$ are measured in the van der Pauw geometry by sourcing at 11 Hz and 17 Hz, respectively.

\bibliography{apssamp}


\end{document}